\documentclass[final,5p,times,authoryear]{elsarticle}

\usepackage{amsmath}
\usepackage{amssymb}
\usepackage{booktabs}
\usepackage{multirow}
\usepackage{array}
\usepackage{graphicx}
\usepackage{subfig}
\usepackage{hyphenat}
\usepackage{orcidlink}
\usepackage{microtype}
\hypersetup{colorlinks=true, linkcolor=blue!60!black, citecolor=blue!60!black, urlcolor=blue!60!black,
            pdftitle={CRUX: Clinical-Text Anchoring for Cross-Institutional CT-Pathology Fusion in Non-Small Cell Lung Cancer Survival Prediction},
            pdfauthor={Kyungwon Kim and Dosik Hwang}}

\journal{Medical Image Analysis}
\myfooter[L]{Preprint}

\begin{document}

\begin{frontmatter}

\title{CRUX: Clinical-Text Anchoring for Cross-Institutional CT--Pathology Fusion in Non-Small Cell Lung Cancer Survival Prediction\tnoteref{t1}}
\tnotetext[t1]{A preliminary version of this work appeared in \citet{kim2024miccai}.}

\author[inst1]{Kyungwon Kim\,\orcidlink{0009-0005-4760-8886}}
\ead{gmkgw93@gmail.com}

\author[inst1,inst2,inst3,inst4]{Dosik Hwang\,\orcidlink{0000-0002-2217-2837}\corref{cor1}}
\ead{dosik.hwang@yonsei.ac.kr}
\cortext[cor1]{Corresponding author}

\affiliation[inst1]{organization={School of Electrical and Electronic Engineering, Yonsei University},
            addressline={50 Yonsei-ro, Seodaemun-gu},
            city={Seoul},
            postcode={03722},
            country={Republic of Korea}}

\affiliation[inst2]{organization={Department of Radiology, College of Dentistry, Yonsei University},
            city={Seoul},
            postcode={03722},
            country={Republic of Korea}}

\affiliation[inst3]{organization={Department of Oral and Maxillofacial Radiology, College of Dentistry, Yonsei University},
            city={Seoul},
            postcode={03722},
            country={Republic of Korea}}

\affiliation[inst4]{organization={Artificial Intelligence and Robotics Institute, Korea Institute of Science and Technology},
            addressline={5 Hwarang-ro 14-gil, Seongbuk-gu},
            city={Seoul},
            postcode={02792},
            country={Republic of Korea}}

\begin{abstract}
Accurate prognosis in non-small cell lung cancer (NSCLC) benefits from integrating computed tomography (CT) and histopathological whole-slide images (WSIs), which capture complementary macroscopic and microscopic disease characteristics. Fusing these modalities is difficult: their features come from different pretrained foundation-model encoders and lack spatial correspondence, and both are sensitive to institution-specific acquisition and staining, causing multimodal models to overfit site-specific artifacts and generalize poorly. We present CRUX (CRoss-modal Unification via teXt anchoring), a fusion framework that uses clinical text---encoded by a biomedical language model---as an institution-invariant semantic anchor. Because clinical variables follow standardized classification systems, their embeddings occupy a representation space that is consistent across institutions, providing a stable reference frame for aligning heterogeneous imaging features. CRUX introduces a text-mediated cross-modal bridge in which CT- and pathology-derived text-query summaries exchange information within a compact text semantic space through a learnable hub token, avoiding direct image-space cross-attention that would propagate domain-specific artifacts. We evaluate on 2{,}157 NSCLC patients from five hospitals, holding out one hospital ($n=446$) entirely for external validation. CRUX attains an external concordance index (C-index) of 0.761, the highest among all unimodal, bimodal, and trimodal configurations. Notably, direct CT--WSI fusion degrades under distribution shift, whereas text anchoring restores and improves cross-institutional generalization. These results indicate that language-model-encoded clinical text is an effective structured prior for robust, generalizable multimodal cancer prognosis.
\end{abstract}


\end{frontmatter}

\section{Introduction}\label{ch3.1}

Lung cancer remains one of the most lethal malignancies worldwide, and accurate prognosis is essential for guiding treatment decisions \citep{siegel2024cancer}. In clinical practice, prognostic assessment draws on two primary imaging modalities: computed tomography (CT), which provides macroscopic characterization of tumor size, location, and morphology for staging and surgical planning, and histopathological whole-slide images (WSIs), which reveal the microscopic cellular architecture that determines definitive diagnosis and histological subtype classification. Both modalities carry complementary prognostic information---CT captures the spatial extent and heterogeneity of disease at the organ level, while WSIs expose the cellular and structural patterns that underlie tumor aggressiveness. Yet, despite their complementary clinical roles, multimodal fusion of CT and WSI for survival prediction has received considerably less attention than the WSI--genomics integration that dominates the computational pathology literature.

A key technical challenge in CT--WSI fusion lies in the heterogeneity of their representation spaces. CT sub-volumes and WSI patches are encoded by fundamentally different pretrained foundation models---3D self-supervised encoders such as FMCIB \citep{pai2024fmcib} for CT and 2D pathology encoders such as UNI \citep{chen2024uni} for WSIs---trained on different data distributions with different architectures, producing feature spaces whose geometric and statistical properties differ substantially. Established cross-attention fusion methods---including MCAT \citep{chen2021mcat}, CMTA \citep{zhou2023cmta}, and SurvPath \citep{jaume2024survpath}---were developed and validated for WSI--genomics integration, where both modalities describe properties of the same tissue specimen. Their effectiveness on heterogeneous image--image combinations such as CT and WSI, which originate from different imaging processes and capture information at vastly different scales, has not been systematically explored. Compounding this, both imaging modalities are sensitive to institution-specific acquisition and preparation, so models tuned on one hospital often fail to transfer to another---a barrier that is rarely quantified in multimodal survival studies.

We address this challenge by leveraging a third source of information---clinical text---as a \emph{semantic anchor} for cross-modal fusion. In existing multimodal survival pipelines, clinical variables are typically reduced to scalar tabular features and processed through simple linear layers or concatenated as auxiliary inputs, limiting their role to low-dimensional side information. We take a fundamentally different approach: structured clinical variables are encoded through biomedical language models such as BioLord \citep{remy2022biolord} and BioBERT \citep{lee2020biobert}, which transform them into dense token-level embeddings that carry population-level prognostic knowledge inherited from large-scale medical-literature pretraining. Crucially, because clinical variables follow standardized classification systems (e.g., AJCC staging, WHO histological classification), these embeddings occupy a representation space that is inherently consistent across institutions---in contrast to imaging features, which are sensitive to scanner characteristics, staining protocols, and institutional practices. This institution-invariant property makes clinical-text embeddings particularly well-suited as a shared semantic reference frame for aligning heterogeneous imaging modalities and for regularizing fusion against domain shift.

Building on a preliminary version of this work \citep{kim2024miccai}, which introduced text-anchored encoding for aligning individual image modalities with clinical text, we propose CRUX (CRoss-modal Unification via teXt anchoring). CRUX extends that foundation in two directions. First, it introduces a \emph{text-mediated cross-modal bridge} that enables indirect CT--WSI interaction through the shared text-anchored representation space, moving beyond the independent per-modality alignment of the preliminary work. Second, it validates the framework through external evaluation on an independent hospital cohort not seen during training, demonstrating that text-anchored fusion generalizes across institutional boundaries---a critical requirement for clinical deployment that is rarely assessed in multimodal survival prediction.

It is important to clarify the role of clinical text. Clinical text does not introduce fundamentally new prognostic information: the recorded variables are themselves derived from image-based assessment and clinical observation. Its role is not to serve as an independent information source but to provide a \emph{structured semantic prior} that establishes a common reference frame for cross-modal alignment and regularizes fusion. This framing is directly supported by our experiments: direct CT--WSI fusion \emph{degrades} under institutional shift, whereas routing cross-modal interaction through the text anchor restores and improves external generalization---even though text is the weakest stand-alone, imaging-free signal.

In summary, our contributions are:
\begin{itemize}
\item A text-anchoring formulation that uses institution-invariant, language-model-encoded clinical text as a shared semantic reference frame for fusing spatially uncorrelated CT and WSI features.
\item A text-mediated cross-modal bridge that routes CT--pathology interaction through a compact text hub with explicit aggregation and broadcast, avoiding artifact-propagating image-space cross-attention---the key extension over our preliminary version \citep{kim2024miccai}.
\item An external-validation study on 2{,}157 non-small cell lung cancer (NSCLC) patients from five hospitals, with one hospital held out entirely, showing that CRUX attains the best external concordance index (C-index, 0.761) among all unimodal, bimodal, and trimodal configurations and that text anchoring is what makes fusion robust to institutional shift.
\end{itemize}

The remainder of the paper is organized as follows. Section~\ref{sec:related} reviews related work. Section~\ref{ch3.2} details the CRUX methodology. Section~\ref{ch3.3} describes the experimental setup, datasets, and evaluation protocol. Section~\ref{ch3.4} presents the main results and ablation studies. Section~\ref{ch3.5} discusses findings and limitations and concludes.

\section{Related Work}\label{sec:related}

\paragraph{Multimodal survival prediction} Combining complementary modalities has repeatedly improved cancer prognosis over unimodal models \citep{baltrusaitis2018multimodal, acosta2022multimodal}. Much of the computational-pathology literature couples WSIs with genomics: early approaches concatenated modality embeddings \citep{cheerla2019deep, mobadersany2018predicting, chen2020porpoise}, while more recent methods introduce explicit cross-modal attention. MCAT \citep{chen2021mcat} uses genomic-guided co-attention over WSI patches; CMTA \citep{zhou2023cmta} adds bidirectional cross-attention with modality-specific and cross-modal streams; and SurvPath \citep{jaume2024survpath} models pathway--patch interactions through pairwise co-attention. These methods assume that the two modalities describe the same specimen and share an alignable structure---an assumption that does not hold for CT and WSI, which are acquired by different processes at vastly different scales.

\paragraph{Foundation-model encoders and MIL aggregation} Both imaging modalities are increasingly represented by frozen foundation models. For pathology, self-supervised encoders such as UNI \citep{chen2024uni}, CTransPath \citep{wang2022ctranspath}, and CONCH \citep{lu2024conch} yield strong patch-level features that are pooled by multiple-instance learning (MIL) \citep{maron1998mil, ilse2018abmil}, with transformer aggregators such as TransMIL \citep{shao2021transmil} and CLAM \citep{lu2021clam} in wide use. For CT, self-supervised 3D encoders such as FMCIB \citep{pai2024fmcib} provide tumor-centric representations. CRUX builds directly on these encoders, keeping them frozen and focusing on how heterogeneous features should be \emph{fused}.

\paragraph{Language models for clinical variables} Biomedical language models such as BioBERT \citep{lee2020biobert} and BioLord \citep{remy2022biolord}, together with vision--language pretraining \citep{radford2021clip}, capture rich medical semantics. In most survival pipelines, however, clinical variables are still consumed as scalar tabular inputs \citep{cox1972regression, friedman2001gbm}. We instead verbalize structured variables and encode them with a biomedical language model, exploiting the institution-invariance of standardized clinical vocabularies as an anchor for imaging fusion.

\paragraph{Domain shift and missing modalities} Histopathology features are notoriously sensitive to staining and scanning differences across institutions, motivating extensive stain-normalization and domain-generalization research. Rather than adapting the feature extractor, CRUX mitigates cross-institutional shift at the \emph{fusion} level by anchoring interaction in a domain-invariant text space. A related but distinct challenge is entirely missing modalities \citep{sharma2019missing, ma2021smil, zhang2022m3care, wang2023shaspec}; CRUX assumes complete modalities and treats robustness to missing inputs as future work.

\section{Methods}\label{ch3.2}
 
\begin{figure*}[t]
\centering
\includegraphics[width=\textwidth]{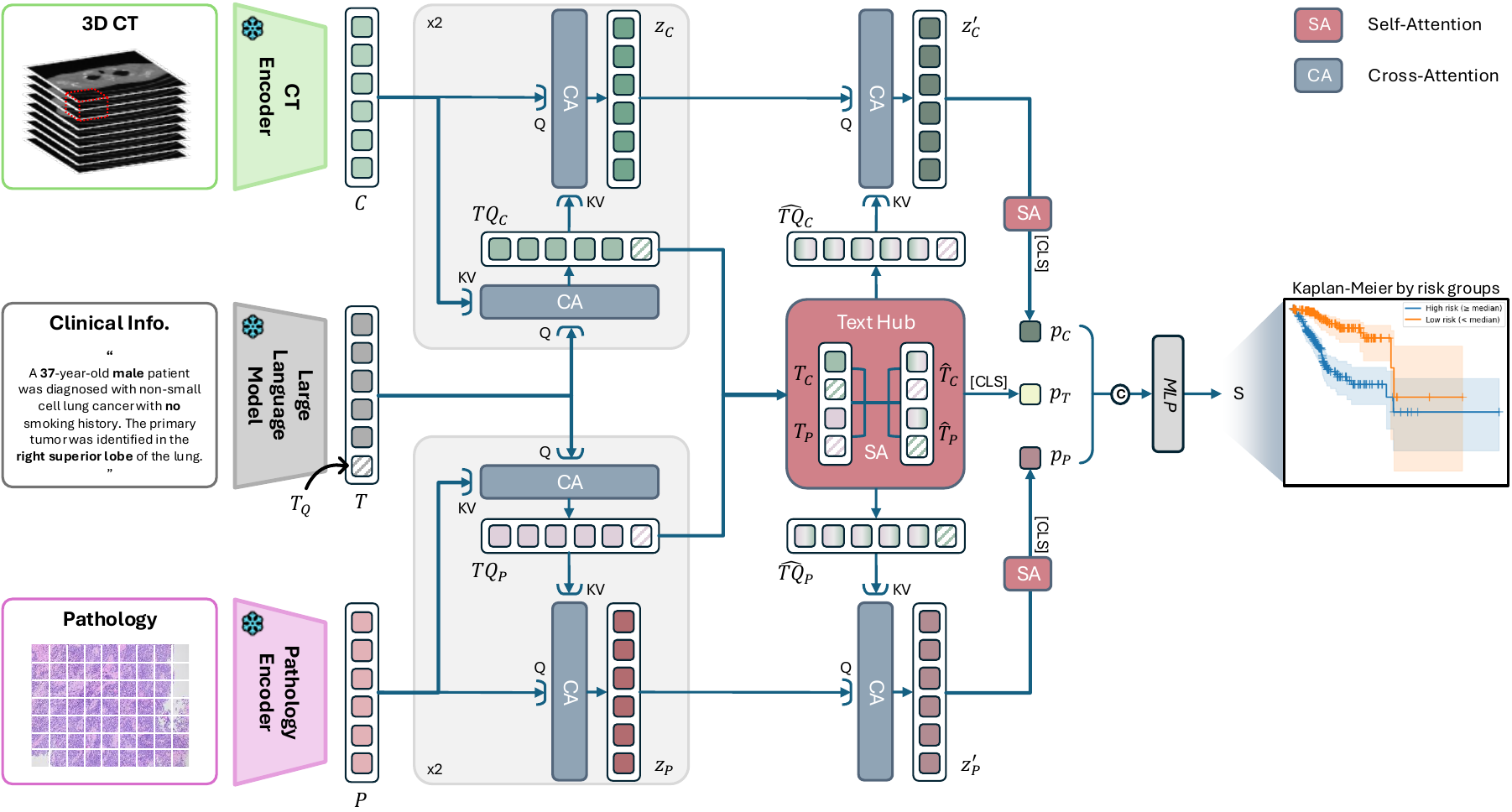}
\caption{Overall architecture of CRUX. The text-anchored encoding module performs iterative bidirectional cross-attention between text-query tokens and each imaging modality, producing enriched text summaries and text-contextualized image representations. The text hub enables cross-modal information exchange in the compact text semantic space and produces a cross-modal consensus representation $\mathbf{p}_T$ via class-token pooling. The enriched summaries are then broadcast to the imaging tokens, and each modality is independently aggregated for survival prediction. The ``clinical text'' input shown here denotes a templated verbalization of structured clinical descriptors (age, sex, smoking history, tumor location), not a free-form clinical, radiology, or pathology note.}
\label{fig:crux_overview}
\end{figure*}

This section describes the CRUX framework for multimodal survival prediction from CT, WSI, and clinical text. An overview of the architecture is shown in Figure~\ref{fig:crux_overview}. The framework consists of three components: text-anchored encoding (Section~\ref{ch3.2.2}), where clinical text tokens and learnable query tokens serve as semantic axes through which imaging features are organized via iterative bidirectional cross-attention; a text-mediated cross-modal bridge (Section~\ref{ch3.2.3}), where the text summaries from both modalities exchange cross-modal information and the resulting enriched summaries are broadcast back to the imaging tokens; and multimodal aggregation followed by survival prediction (Section~\ref{ch3.2.4}).
 
\subsection{Feature Extraction}\label{ch3.2.1}
 
Each modality is encoded by a domain-specific pretrained model and projected into a common $D$-dimensional embedding space ($D = 512$).
 
\paragraph{CT} Preoperative CT volumes are encoded using FMCIB~\citep{pai2024fmcib}, a foundation model pretrained on a large corpus of medical imaging data via self-supervised learning. A tumor-centered sub-volume of $50 \times 50 \times 50$ voxels (resampled to $1.0\,\text{mm}$ isotropic spacing) is cropped from each scan and passed through the 3D ResNet backbone, which produces a feature tensor of shape $4096 \times 6 \times 6 \times 6$ corresponding to the spatial grid after successive downsampling. This tensor is reshaped into $N_C = 216$ spatial tokens of dimension $4096$. A two-layer MLP projects these tokens to the shared embedding space, yielding $\mathbf{C} \in \mathbb{R}^{N_C \times D}$.
 
\paragraph{Pathology} Gigapixel WSIs are partitioned into non-overlapping patches at diagnostic magnification, and each patch is encoded using UNI~\citep{chen2024uni}, a vision transformer pretrained on over 100 million histopathology images. This produces $N_P$ patch-level features of dimension $1536$, where $N_P$ varies per slide (typically $500$--$4000$). A linear layer with dropout projects these features to $\mathbf{P} \in \mathbb{R}^{N_P \times D}$.
 
\paragraph{Clinical text} The meaning of ``text'' in this framework warrants clarification at the outset, as the term can be misleading in a clinical context. The text modality is \emph{not} a free-form clinical note, radiology report, or pathology report; it is a templated natural-language representation of structured, routinely available clinical descriptors (a ``language-encoded clinical descriptor''). Specifically, a small set of standardized categorical and numerical variables is verbalized into a fixed-template sentence and encoded by a biomedical language model. No free-text electronic medical record content is used. A key design decision is the exclusion of AJCC TNM staging variables and histological subtype from the text input. TNM staging (overall stage, T, N, M) is derived from expert interpretation of CT images, while histological subtype is determined by pathological examination of WSIs---both represent the \emph{output} of imaging analysis, not independent sources of information. Including these variables would provide the model with expert-derived conclusions that render the imaging encoders redundant. Instead, the text modality is restricted to pre-diagnostic patient information: demographics (age, sex, smoking history) and anatomical tumor location, which is known prior to imaging interpretation as a prerequisite for tissue acquisition.
 
These variables are composed into a natural language paragraph using a template:
 
\begin{quote}
\small\textit{``A \textbf{65}-year-old \textbf{male} patient was diagnosed with non-small cell lung cancer with \textbf{no} smoking history. The primary tumor was identified in the \textbf{right superior lobe} of the lung.''}
\end{quote}
 
\noindent where the bold segments indicate patient-specific values populated from the clinical record. The paragraph is encoded using BioLord~\citep{remy2022biolord}, a biomedical sentence embedding model trained on clinical ontologies, producing $N_T$ token-level embeddings of dimension $768$. A two-layer MLP projects these to $\mathbf{T} \in \mathbb{R}^{N_T \times D}$.

\subsection{Text-Anchored Encoding}\label{ch3.2.2}
 
The text-anchored encoding module can be understood as a \emph{low-rank, text-guided self-attention} over the imaging tokens. In standard self-attention, each imaging token attends to all other tokens, forming an $N_X \times N_X$ attention matrix. In contrast, this module routes all interaction through a small set of text tokens that serve as semantic axes: each text token first summarizes the imaging features relevant to its clinical concept, and then each imaging token retrieves information from these summaries. The result is equivalent to self-attention constrained to a rank determined by the number of text tokens, where the axes correspond to clinically meaningful concepts such as patient demographics and tumor location.

\paragraph{Learnable query tokens} The $N_T$ patient-specific text tokens provide domain-invariant anchoring axes, but they are limited to the clinical concepts present in the input paragraph. To expand the representational capacity, $N_Q$ learnable query tokens $\mathbf{Q} \in \mathbb{R}^{N_Q \times D}$ are introduced as additional axes that are free to discover survival-relevant imaging patterns directly from the training signal. Both the text and query tokens are projected through modality-specific linear layers before entering the encoding module:
\begin{equation}\label{eq:crux_text_proj}
\begin{aligned}
    \mathbf{TQ}^C &= [f^T_C(\mathbf{T});\; f^Q_C(\mathbf{Q})], \\
    \mathbf{TQ}^P &= [f^T_P(\mathbf{T});\; f^Q_P(\mathbf{Q})]
\end{aligned}
\end{equation}
where $f^T_C, f^Q_C, f^T_P, f^Q_P$ are learned projections. The per-modality projection allows the same clinical concept or query to take on different roles depending on the imaging modality---for example, a query token projected for CT may learn to attend to tumor boundary regions, while the same query projected for pathology may attend to cellular morphology patterns. The augmented token sequence $\mathbf{TQ}^X \in \mathbb{R}^{(N_T + N_Q) \times D}$ serves as the set of semantic axes for each modality.
 
\paragraph{Bidirectional cross-attention} For a given imaging modality $\mathbf{X} \in \{\mathbf{C}, \mathbf{P}\}$ and its corresponding augmented tokens $\mathbf{TQ}^X$, the encoding module performs $L$ iterations ($L=2$ in this work) of bidirectional cross-attention. Each iteration consists of three operations---text-to-image cross-attention, MLP refinement, and image-to-text cross-attention---with residual connections and layer normalization:
\begin{equation}\label{eq:crux_bca}
\begin{split}
    \mathbf{TQ}^X &\leftarrow \text{LN}\!\left(\mathbf{TQ}^X + \text{CrossAttn}(\mathbf{TQ}^X, \mathbf{X})\right), \\
    \mathbf{TQ}^X &\leftarrow \text{LN}\!\left(\mathbf{TQ}^X + \text{MLP}(\mathbf{TQ}^X)\right)
\end{split}
\end{equation}
\begin{equation}\label{eq:crux_bca2}
    \mathbf{X} \leftarrow \text{LN}\!\left(\mathbf{X} + \text{CrossAttn}(\mathbf{X}, \mathbf{TQ}^X)\right)
\end{equation}
In the first pass, each text-query token aggregates visual evidence from the imaging tokens relevant to its semantic concept (Equation~\ref{eq:crux_bca}). In the second pass, each imaging token is re-contextualized by attending to the enriched text-query summaries (Equation~\ref{eq:crux_bca2}). Repeating this process over $L$ layers enables iterative refinement, where each layer operates on increasingly aligned representations.

After $L$ iterations, the module produces two outputs per modality: a text-query summary $\mathbf{TQ}_X \in \mathbb{R}^{(N_T + N_Q) \times D}$ capturing modality-specific visual evidence organized along the semantic axes, and a text-contextualized image representation $\mathbf{z}_X$. Applying this to both imaging modalities yields $\mathbf{TQ}_C$, $\mathbf{TQ}_P$, $\mathbf{z}_C \in \mathbb{R}^{N_C \times D}$, and $\mathbf{z}_P \in \mathbb{R}^{N_P \times D}$.

\subsection{Text-Mediated Cross-Modal Bridge}\label{ch3.2.3}
 
CT and pathology capture complementary aspects of cancer---macroscopic morphology and microscopic cellular architecture, respectively---but their feature representations occupy distinct spaces shaped by different pretrained encoders. Direct cross-attention between CT and pathology tokens is computationally expensive ($N_C \times N_P$ attention matrix) and semantically challenging given the representational heterogeneity. CRUX addresses this by mediating all cross-modal interaction through the text semantic space.
 
\paragraph{Text hub} The text-query summaries $\mathbf{TQ}_C$ and $\mathbf{TQ}_P$ from the text-anchored encoding carry visual evidence from their respective imaging modalities, but each summary only knows about its own modality. To enable cross-modal information exchange, these summaries are concatenated along with a prepended learnable class token $\mathbf{c} \in \mathbb{R}^{D}$ and processed by self-attention:
\begin{equation}\label{eq:crux_hub}
    [\mathbf{p}_T;\; \hat{\mathbf{TQ}}_C;\; \hat{\mathbf{TQ}}_P] = \text{SelfAttn}\!\left([\mathbf{c};\; \mathbf{TQ}_C;\; \mathbf{TQ}_P]\right)
\end{equation}
where $\mathbf{p}_T \in \mathbb{R}^{D}$ is the output corresponding to the class token and $[{\cdot}\,;\,{\cdot}]$ denotes concatenation along the token dimension. The self-attention operates over the full $1 + 2(N_T + N_Q)$ tokens, allowing each text-query token enriched with CT evidence to attend to tokens enriched with pathology evidence, and vice versa. This is the core contribution of CRUX over the prior text-anchored framework~\citep{kim2024miccai}: whereas the prior work processes each imaging modality independently through text, CRUX introduces an explicit cross-modal exchange mechanism that operates entirely within the compact text semantic space ($2(N_T + N_Q) \ll N_C + N_P$). The class token output $\mathbf{p}_T$ serves as a cross-modal consensus representation that summarizes the joint information from both imaging modalities as interpreted through the clinical text axes. The remaining tokens are split back into modality-specific views $\hat{\mathbf{TQ}}_C, \hat{\mathbf{TQ}}_P \in \mathbb{R}^{(N_T + N_Q) \times D}$, each now carrying information from both imaging modalities.
 
\paragraph{Cross-modal broadcast} The cross-modal enriched text-query summaries are delivered to the imaging tokens through cross-attention followed by an MLP:
\begin{equation}\label{eq:crux_broadcast}
\begin{aligned}
    \mathbf{z}'_C &= \text{Broadcast}(\mathbf{z}_C, \hat{\mathbf{TQ}}_C), \\
    \mathbf{z}'_P &= \text{Broadcast}(\mathbf{z}_P, \hat{\mathbf{TQ}}_P)
\end{aligned}
\end{equation}
where $\text{Broadcast}(\mathbf{z}, \hat{\mathbf{TQ}})$ consists of a residual cross-attention ($\mathbf{z}$ attends to $\hat{\mathbf{TQ}}$) followed by an MLP with a residual connection. The residual design preserves the text-contextualized image representation from the encoding module while adding cross-modal information from the hub. The enriched representations $\mathbf{z}'_C$ and $\mathbf{z}'_P$ are then aggregated by modality-specific self-attention encoders with class-token pooling to produce pooled representations $\mathbf{p}_C$ and $\mathbf{p}_P$.

\subsection{Prediction and Training Objective}\label{ch3.2.4}
 
\paragraph{Survival prediction} The final prediction integrates three complementary representations: the imaging-derived representations $\mathbf{p}_C$ and $\mathbf{p}_P$, which capture individual-level visual detail from each modality, and the cross-modal consensus $\mathbf{p}_T$ from the text hub, which summarizes the joint clinical interpretation of both imaging modalities through the text semantic axes:
\begin{equation}\label{eq:crux_final}
    h_{\text{score}} = \phi\!\left([\mathbf{p}_C;\; \mathbf{p}_P;\; \mathbf{p}_T]\right)
\end{equation}
where $h_{\text{score}} \in \mathbb{R}^{n_{\text{bins}}}$ represents the predicted per-bin hazard logits and $\phi$ is a two-layer MLP with ReLU activation and dropout. Notably, $\mathbf{p}_T$ is not a prediction from clinical text alone; it is the class-token output of a self-attention encoder that has jointly processed text-query summaries from both CT and pathology. Its role in the final prediction is to provide a stable, domain-invariant summary that complements the potentially noisy but informationally rich imaging representations.
 
\paragraph{Training objective} The model is trained with the discrete-time negative log-likelihood survival loss $\mathcal{L}_{\text{surv}}$ \citep{zadeh2020nll}. In addition to the main prediction from the concatenated representation (Equation~\ref{eq:crux_final}), auxiliary survival losses are computed independently from each of the three pooled representations $\mathbf{p}_C$, $\mathbf{p}_P$, and $\mathbf{p}_T$ through separate linear heads, weighted by $\alpha_{\text{aux}} = 0.1$. These auxiliary objectives encourage each representation stream to maintain independent prognostic capacity, preventing the model from relying exclusively on a single modality.

\section{Experiments}\label{ch3.3}
 
\subsection{Datasets}\label{ch3.3.1}
 
\paragraph{Data source and inclusion criteria}
This study uses lung cancer prognosis prediction data from the Open AI Dataset Project (AI-Hub, South Korea), a publicly accessible multi-institutional dataset collected across university hospitals. From the original collection, patients meeting the following criteria were included: (1) histologically confirmed adenocarcinoma or squamous cell carcinoma, which together constitute the majority of non-small cell lung cancer cases; (2) surgical resection performed as primary treatment; (3) availability of paired pretreatment CT, histopathological WSI, and structured clinical information; and (4) valid clinical variable entries across all staging, demographic, and histological fields. Patients whose death was attributed to non-cancer causes were excluded to ensure that the survival endpoint reflects cancer-specific prognosis. Additionally, a small number of cases with data quality issues---such as CT--pathology identifier mismatches or corrupted imaging data---were removed after manual inspection. No restriction was imposed on minimum follow-up duration; patients who were alive at last contact or lost to follow-up were treated as right-censored observations in accordance with standard survival analysis conventions. After applying these criteria across five hospitals, the final cohort comprised 2,157 patients with complete multimodal data. Table~\ref{tab:crux_cohort} summarizes the cohort composition.
 
\begin{table}[t]
\centering
\caption{Cohort composition across five hospitals. The development cohort is used for 5-fold cross-validation, while PNUH is held out entirely as the external validation cohort.}
\label{tab:crux_cohort}
\resizebox{\columnwidth}{!}{\begin{tabular}{llccc}
\toprule
Cohort & Hospital & $N$ & Event (\%) & Role \\
\midrule
\multirow{4}{*}{Development} & AJMC  & 623 & \multirow{4}{*}{6.5} & \multirow{4}{*}{5-fold CV} \\
                             & CNUH  & 689 & & \\
                             & HUMC  & 219 & & \\
                             & SCHMC & 180 & & \\
\cmidrule{2-3}
                             & \textit{Subtotal} & \textit{1,711} & & \\
\midrule
External validation          & PNUH  & 446 & 13.0 & Held-out test \\
\midrule
\multicolumn{2}{l}{\textbf{Total}} & \textbf{2,157} & & \\
\bottomrule
\end{tabular}
}
\end{table}
 
\paragraph{Development and external validation cohorts}
To assess the generalizability of the proposed framework, the dataset is divided into a \emph{development cohort} used for model training and cross-validation, and an \emph{external validation cohort} held out entirely during training.
 
The selection of the external validation cohort was guided by a systematic analysis of inter-institutional distribution differences. For each candidate hospital (and pairwise combinations), the distributions of ten clinical variables---including demographics (age, sex, smoking history), staging (overall TNM, T, N, M), tumor location, histological subtype, and survival outcome---were compared between the candidate external cohort and the remaining development cohort using chi-squared or Mann-Whitney U tests as appropriate.
 
PNUH was selected as the external validation cohort based on three considerations. First, it provides sufficient sample size ($N = 446$, 20.7\% of the total cohort) to ensure statistically reliable evaluation. Second, demographic variables (age, sex, smoking history) and histological subtype show no significant differences between the development and external cohorts ($p > 0.05$), ensuring that any observed performance gap reflects genuine distributional shift in disease characteristics rather than population-level confounding. Third, the two cohorts exhibit statistically significant differences in staging-related variables (overall TNM stage, T, and N; all $p < 0.05$) and in event rate (13.0\% vs.\ 6.5\%, $p < 0.001$), indicating that PNUH presents a more severe case mix with higher cancer-related mortality. This pattern represents a clinically realistic and challenging generalization scenario: the external cohort shares the same demographic background but presents a different distribution of disease severity and outcomes, mirroring the variation that would be encountered when deploying a model to a new hospital.
 
\begin{figure*}[t]
\centering
\includegraphics[width=\textwidth]{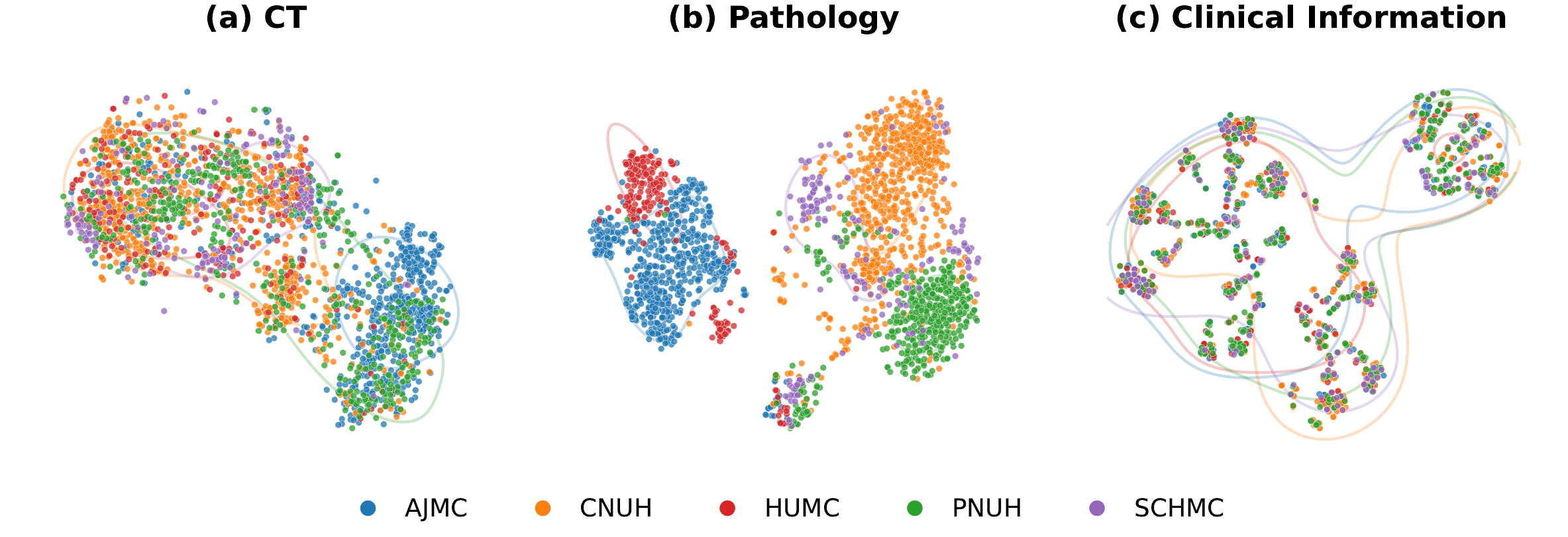}
\caption{t-SNE visualization of per-modality feature distributions across five hospitals. (a) CT features (FMCIB) show moderate hospital-specific clustering. (b) Pathology features (UNI) exhibit pronounced institutional separation. (c) Clinical text embeddings (BioLord) overlap substantially across hospitals, reflecting the standardized nature of clinical classification systems.}
\label{fig:crux_tsne}
\end{figure*}

\paragraph{Feature distribution analysis}
Figure~\ref{fig:crux_tsne} visualizes the per-modality feature distributions across hospitals using t-SNE. The three modalities exhibit markedly different inter-institutional characteristics. CT features (FMCIB) show moderate hospital-specific clustering with partial overlap. Pathology features (UNI) exhibit pronounced hospital-specific separation, with each institution forming a distinct cluster---reflecting known sensitivity of histopathological features to staining protocols, slide preparation, and scanning equipment. In contrast, clinical text embeddings (BioLord) show substantial overlap across all hospitals, consistent with the standardized nature of the clinical classification systems (AJCC staging, WHO histological classification) that underlie these variables. This observation provides a representational basis for the text-anchored fusion strategy proposed in this work: text embeddings occupy a representation space that is inherently more consistent across institutions than either imaging modality.
 
\subsection{Implementation Details}\label{ch3.3.2}
 
\paragraph{Feature extraction} All pretrained encoders are frozen and used solely for offline feature extraction. CT features are extracted using FMCIB~\citep{pai2024fmcib} from tumor-centered $50 \times 50 \times 50$ sub-volumes, yielding $216$ tokens of dimension $4096$. Pathology features are extracted using UNI~\citep{chen2024uni} from non-overlapping $224 \times 224$ patches at $10\times$ magnification, with a maximum of $2048$ patches per slide to limit memory consumption. Clinical text features are extracted per token using BioLord~\citep{remy2022biolord} from the paragraph template described in Section~\ref{ch3.2.1}, producing $N_T$ tokens of dimension $768$ where $N_T$ depends on the tokenizer output for each patient's paragraph (typically $30$--$40$ tokens).
 
\paragraph{Training} All models share a common embedding dimension of $D = 512$ and are trained with the AdamW optimizer ($\beta_1 = 0.9$, $\beta_2 = 0.999$, weight decay $5 \times 10^{-4}$) using a cosine annealing learning rate schedule with $5$ epochs of linear warmup, a peak learning rate of $5 \times 10^{-5}$, and a minimum learning rate of $1 \times 10^{-6}$. Training proceeds for a maximum of $30$ epochs with early stopping triggered after $10$ consecutive validations without improvement. Gradient clipping (max norm $1.0$) is applied. Due to the variable sequence lengths across patients, all models are trained with batch size $1$ on a single NVIDIA A6000 GPU per fold. For methods employing text-anchored encoding, the bidirectional cross-attention uses $L = 2$ layers with $8$ attention heads and an MLP hidden dimension of $2048$. CRUX additionally uses $N_Q = 4$ learnable query tokens and an auxiliary loss weight of $\alpha_{\text{aux}} = 0.1$ applied to the per-modality and cross-modal consensus predictions. All self-attention encoders used for token aggregation follow the Nystr\"{o}m attention architecture~\citep{xiong2021nystromformer} with positional encoding adapted to the spatial structure of each modality (3D grid for CT, 2D grid for pathology).
 
\paragraph{Cross-validation} The development cohort ($N = 1{,}711$) is divided into $5$ folds using censorship\hyp{}stratified splitting: within each hospital, patients are stratified by their censorship status (event vs.\ censored) to ensure that the low event rate ($6.5\%$) is approximately balanced across folds. For each fold, four folds serve as training data and one fold as the validation set. Model selection is based on the best validation concordance index (C-index). The held-out external validation cohort (PNUH, $N = 446$) is evaluated using each fold's best checkpoint independently, and the results are reported as the mean and standard deviation across folds.
 
\subsection{Baseline Methods}\label{ch3.3.3}
 
To evaluate the contributions of multimodal fusion and text-anchored interaction, this study compares a comprehensive set of baselines spanning unimodal, bimodal, and trimodal configurations. All baseline methods share the same pretrained feature extractors, projection layers, and training protocol described in Section~\ref{ch3.3.2}, differing only in the fusion architecture.
 
\paragraph{Unimodal baselines} Each modality is evaluated independently to establish per-modality performance bounds. For CT and pathology, extracted features are aggregated via self-attention with class-token pooling and passed through a linear prediction head. For clinical text, two settings are considered: (1) the BioLord token embeddings aggregated via self-attention, and (2) traditional machine learning methods---Cox proportional hazards (CoxPH)~\citep{cox1972regression} and gradient boosted machines (GBM)~\citep{friedman2001gbm}---applied to scalar-valued clinical features. The comparison between deep learning and traditional approaches on the same clinical variables isolates the contribution of the pretrained text encoder's semantic knowledge.
 
\paragraph{Bimodal baselines} Bimodal fusion is evaluated for all three modality pairs (CT+pathology, CT+text, pathology+text) using the following architectures. \textbf{Concatenation} serves as the simplest baseline, where features from both modalities are aggregated independently, concatenated, and passed through a joint prediction head. \textbf{MCAT}~\citep{chen2021mcat} applies genomic-guided cross-attention where one modality attends to the other; in this setting, the smaller modality (text or CT) serves as the query attending to the larger modality. \textbf{SurvPath}~\citep{jaume2024survpath} employs pairwise co-attention where the smaller modality attends to the concatenation of both modalities, while the larger modality attends to the smaller one. \textbf{CMTA}~\citep{zhou2023cmta} applies bidirectional cross-attention between the two modalities followed by modality-specific aggregation, producing both direct and cross-modal representations that are combined for prediction.
 
\paragraph{Trimodal baselines} Methods integrating all three modalities are compared under a unified evaluation:
\begin{itemize}
    \item \textbf{Concatenation}: each modality is independently aggregated, the three pooled representations are concatenated, and a joint prediction head produces the final output. This serves as the simplest feature-level fusion baseline.
    \item \textbf{Text-anchored bag}~\citep{kim2024miccai}: the prior text-anchored framework, where bidirectional cross-attention between text and each imaging modality produces text-contextualized image tokens and image-enriched text tokens. All four token groups are concatenated into a single multimodal bag and jointly aggregated by a shared self-attention encoder.
    \item \textbf{SurvPath (trimodal)}~\citep{jaume2024survpath}: extended to three modalities by applying pairwise co-attention between all modality pairs (CT--pathology, pathology--text, text--CT), with each pair producing pooled representations that are concatenated for prediction.
    \item \textbf{CRUX}: the proposed method (Section~\ref{ch3.2}).
\end{itemize}
 
\subsection{Evaluation Protocol}\label{ch3.3.4}
 
Model performance is assessed using the concordance index (C-index), which measures the probability that, for a randomly selected pair of patients, the model assigns a higher risk score to the patient who experiences the event first. The C-index is computed separately on the development cohort (mean across 5-fold cross-validation) and on the external validation cohort (PNUH), with standard deviations reported across folds in both cases. For external evaluation, predictions from the five fold-trained models are averaged to produce a single ensemble prediction per patient, yielding a single C-index value without standard deviation.
 
The primary evaluation criterion is external validation performance, as it reflects the model's ability to generalize to an unseen hospital with different disease severity distributions. The gap between development and external performance is reported as a relative indicator of generalization robustness across methods, noting that the two cohorts differ in disease severity and event rate, so the gap should not be interpreted as an absolute measure of generalization quality.
 
Statistical significance of risk stratification is assessed using the log-rank test on Kaplan-Meier survival curves, where patients are divided into high-risk and low-risk groups based on the median predicted risk score. This evaluation is performed on the external validation cohort to assess whether the model's risk predictions translate to meaningful survival differences in an independent population.

\section{Results}\label{ch3.4}
 
\subsection{Main Results}\label{ch3.4.1}
 
Table~\ref{tab:crux_main} presents the comparison across unimodal, bimodal, and trimodal configurations. All methods are evaluated on both the development cohort (5-fold cross-validation) and the external validation cohort (PNUH).
 
\begin{table*}[t]
\centering
\caption{C-index comparison across unimodal, bimodal, and trimodal configurations. C: CT, P: pathology, T: clinical text. Internal: mean $\pm$ standard deviation across 5-fold cross-validation on the development cohort. External: mean prediction from 5-fold trained models (fold ensemble) on the held-out PNUH cohort. The best external result within each modality group is \textbf{bolded}.}
\label{tab:crux_main}
\begin{tabular}{cl ccc cc}
\toprule
Category & Method & C & P & T & Internal & External \\
\midrule
\multirow{3}{*}{Unimodal}
 & Self-attention   & $\checkmark$ &              &              & $0.607 \pm 0.032$ & $0.584$ \\
 & Self-attention   &              & $\checkmark$ &              & $0.777 \pm 0.037$ & $0.616$ \\
 & Self-attention   &              &              & $\checkmark$ & $0.714 \pm 0.025$ & $\mathbf{0.732}$ \\
\midrule
\multirow{12}{*}{Bimodal}
 & Concat           & $\checkmark$ & $\checkmark$ &              & $0.788 \pm 0.033$ & $0.552$ \\
 & MCAT             & $\checkmark$ & $\checkmark$ &              & $0.745 \pm 0.051$ & $\mathbf{0.608}$ \\
 & SurvPath         & $\checkmark$ & $\checkmark$ &              & $0.777 \pm 0.040$ & $0.575$ \\
 & CMTA             & $\checkmark$ & $\checkmark$ &              & $0.739 \pm 0.052$ & $0.601$ \\
\cmidrule{2-7}
 & Concat           & $\checkmark$ &              & $\checkmark$ & $0.759 \pm 0.029$ & $0.746$ \\
 & MCAT             & $\checkmark$ &              & $\checkmark$ & $0.765 \pm 0.023$ & $0.750$ \\
 & SurvPath         & $\checkmark$ &              & $\checkmark$ & $0.753 \pm 0.029$ & $0.755$ \\
 & CMTA             & $\checkmark$ &              & $\checkmark$ & $0.745 \pm 0.023$ & $\mathbf{0.757}$ \\
\cmidrule{2-7}
 & Concat           &              & $\checkmark$ & $\checkmark$ & $0.788 \pm 0.044$ & $0.686$ \\
 & MCAT             &              & $\checkmark$ & $\checkmark$ & $0.782 \pm 0.058$ & $0.655$ \\
 & SurvPath         &              & $\checkmark$ & $\checkmark$ & $0.789 \pm 0.054$ & $0.724$ \\
 & CMTA             &              & $\checkmark$ & $\checkmark$ & $0.772 \pm 0.055$ & $\mathbf{0.758}$ \\
\midrule
\multirow{4}{*}{Trimodal}
 & Concat               & $\checkmark$ & $\checkmark$ & $\checkmark$ & $0.792 \pm 0.041$ & $0.749$ \\
 & Text-anchored bag    & $\checkmark$ & $\checkmark$ & $\checkmark$ & $0.804 \pm 0.044$ & $0.711$ \\
 & SurvPath-Tri         & $\checkmark$ & $\checkmark$ & $\checkmark$ & $0.806 \pm 0.046$ & $0.698$ \\
 & \textbf{CRUX (proposed)} & $\checkmark$ & $\checkmark$ & $\checkmark$ & $0.819 \pm 0.041$ & $\mathbf{0.761}$ \\
\bottomrule
\end{tabular}%

\end{table*}
 
The results validate the core design premises of CRUX. Language model--encoded clinical text provides a stable semantic anchor for cross-institutional generalization, text-mediated fusion systematically improves robustness compared to direct image-only interaction, and the proposed text hub mechanism achieves the highest external C-index among all configurations. The following subsections present these findings organized around the design hypotheses of the proposed framework.

\subsection{Effectiveness of Language Model--Encoded Text}\label{ch3.4.2}
 
A central premise of CRUX is that clinical text, when encoded through biomedical language models, provides a semantically rich and institution-invariant representation that can serve as an anchoring reference for cross-modal fusion. The unimodal and bimodal results provide strong evidence for this premise.
 
\paragraph{Language-model encoding, not added variables or model capacity, drives the gain} The text-only model using BioLord embeddings achieves an external C-index of 0.732 from only four non-staging variables (age, sex, smoking history, location). As shown in Table~\ref{tab:crux_ml}, trained models applied to the same four variables as raw scalars achieve substantially lower performance (external 0.684/0.689/0.638 for CoxPH/GBM/MLP); the same four variables thus yield at most 0.689 as raw scalars but 0.732 when language-encoded. Three observations show that this gain stems from the representation rather than from added information, more variables, or a more flexible model. \emph{First, it is not added variables}: smoking and location are individually weak (external 0.613 for smoking, and tumor location is a nominal lobe label with essentially no transferable ordering, $\approx 0.557$), and adding them to age and sex barely changes the scalar CoxPH (0.680 to 0.684, with an internal decrease), whereas under language encoding the same additions lift BioLord-based TransMIL from 0.658/0.662 (age+sex) to 0.714/0.732 (all four)---a gain of $+0.070$ externally. \emph{Second, it is not model capacity}: a tabular MLP on the same four scalars reaches only 0.638, well below the language-encoded model, so the improvement is not attributable to deep learning alone. \emph{Third, the encoding injects pretrained medical knowledge}: language-model encoding captures population-level knowledge that scalar features cannot access, closing a substantial portion of the gap between non-staging and full-staging predictions ($+0.048$ over CoxPH with the same four variables). The remaining gap to staging-inclusive CoxPH ($-0.036$; 0.732 vs.\ 0.768) reflects the genuine prognostic information in staging variables, which the model must instead extract from imaging---validating the decision to exclude staging from the text input.

\paragraph{Age and sex are informative but insufficient for ranking}
A clinical observation raised in review is that age and sex alone already convey substantial prognostic information. This is borne out, and the directions recovered from the data agree with established clinical knowledge: ranking patients purely by age (no model fitting) yields an external C-index of 0.657 (older patients at higher risk), and by sex 0.600 (male patients at higher risk; the external event rate is $16.2\%$ in men versus $7.1\%$ in women). These single variables are therefore strong, clinically valid hints, but they are insufficient for precise risk ranking: age and sex jointly reach only 0.680, and all four non-staging variables 0.684--0.689---far below staging-inclusive CoxPH (0.768) and multimodal CRUX (0.761). Two clarifications follow. First, survival discrimination is measured by the C-index, a censoring-aware ranking metric rather than classification accuracy; a C-index near 0.68 is clearly above chance (0.5) but far from a deployable ceiling, so an intuition phrased as high ``accuracy'' does not transfer to the ranking task. Second, a single variable's natural ordering (age, 0.657) exceeds a flexible model trained on all four raw scalars only when that model is under-regularized; with proper regularization the trained models recover the expected ordering (GBM 0.689, above age alone), confirming that the limitation is the weak, distribution-shift-prone signal in raw tabular variables rather than any individual feature being harmful---the limitation that language encoding overcomes.

This transfer-stability advantage also explains an apparent paradox with the imaging modalities and directly motivates text anchoring. Internally, imaging carries far more prognostic information than the clinical variables (Table~\ref{tab:crux_main}: pathology $0.777$ and CT+pathology $0.745$, versus $0.638$ for the four raw clinical variables); yet imaging is highly institution-specific and collapses under transfer (pathology from $0.777$ to $0.616$, a $21\%$ drop), so \emph{externally} the modest but domain-invariant clinical signal ($0.689$) exceeds every image-only configuration (CT $0.584$, pathology $0.616$, CT+pathology $0.608$). The point is not that the clinical variables contain more information---internally they do not---but that they transfer stably while information-rich imaging does not. Naively fusing the two images (CT+pathology) does not resolve this, as it only combines two domain-fragile signals; CRUX instead routes the interaction through the stable clinical/text anchor, which restores the value of imaging under transfer and surpasses both regimes (CRUX $0.761$ vs.\ clinical-only $0.689$/$0.732$ and image-only $\leq 0.616$).
 
\begin{table*}[t]
\centering
\caption{Clinical-variable survival baselines (C-index). Internal: development cohort; External: PNUH (5-fold ensemble). Non-staging: age, sex, smoking, location; full: + TNM staging + histological subtype.}
\label{tab:crux_ml}
\begin{tabular}{llcc}
\toprule
Method & Input & Internal & External \\
\midrule
\multicolumn{4}{l}{\textit{Single variable (training-free concordance)}} \\
\quad Age (higher risk with age)  & 1 variable & 0.606 & 0.657 \\
\quad Sex (higher risk in males)   & 1 variable & 0.599 & 0.600 \\
\quad Smoking (higher risk with smoking)   & 1 variable & 0.568 & 0.613 \\
\quad Location (nominal; random)   & 1 variable & 0.510 & 0.557 \\
\midrule
\multicolumn{4}{l}{\textit{Trained models --- raw clinical variables}} \\
\quad CoxPH              & age + sex        & 0.652 & 0.680 \\
\quad CoxPH              & non-staging 4    & 0.638 & 0.684 \\
\quad GBM                & non-staging 4    & 0.622 & 0.689 \\
\quad Tabular MLP (DL)   & non-staging 4    & 0.535 & 0.638 \\
\midrule
\multicolumn{4}{l}{\textit{Language-encoded clinical descriptors}} \\
\quad TransMIL (BioLord) & age + sex        & 0.658 & 0.662 \\
\quad TransMIL (BioLord) & non-staging 4    & \textbf{0.714} & \textbf{0.732} \\
\midrule
\multicolumn{4}{l}{\textit{Trained models --- with TNM staging \& subtype}} \\
\quad CoxPH              & full 9           & 0.767 & 0.768 \\
\quad GBM                & full 9           & 0.743 & 0.772 \\
\bottomrule
\end{tabular}

\end{table*}
 
\paragraph{Text anchoring improves cross-institutional robustness} The value of text as a fusion anchor is evident from the bimodal comparisons. Imaging-only fusion (C+P) achieves external C-indices of 0.552--0.608, consistently below the text-only baseline of 0.732. In contrast, all CT+Text configurations achieve 0.746--0.757, and Pathology+Text configurations reach up to 0.758 (P+T CMTA). This systematic pattern---that text inclusion improves external generalization regardless of fusion architecture---supports the design choice of incorporating text as a semantic anchor rather than relying on imaging modalities alone.
 
\subsection{Text-Mediated Fusion vs.\ Direct Image Fusion}\label{ch3.4.3}
 
The bimodal results illuminate why text-mediated cross-modal interaction is preferable to direct image--image fusion in the multi-institutional setting.
 
\paragraph{Direct image fusion degrades under domain shift} All CT+Pathology fusion methods---Concat, MCAT, SurvPath, and CMTA---exhibit external C-indices of 0.552--0.608, consistently below the text-only baseline (0.732). These methods perform direct cross-attention between CT and pathology tokens, allowing institution-specific artifacts in one modality to propagate to the other. The absence of a domain-invariant reference frame means that the cross-attention patterns learned on the development cohort do not transfer reliably to the external cohort.
 
\paragraph{Text-mediated fusion preserves generalization} When text is included as a fusion partner, the pattern changes qualitatively. CT+Text models achieve 0.746--0.757 externally, and Pathology+Text models achieve 0.655--0.758 with reduced gaps compared to the imaging-only counterparts. This consistent improvement across all fusion architectures (Concat, MCAT, SurvPath, CMTA) confirms that the benefit derives from the text modality itself---specifically its institution-invariant embedding space---rather than from any particular fusion mechanism. This finding motivates the CRUX design: rather than allowing CT and pathology to interact directly, all cross-modal communication is routed through the text semantic space.
 
\subsection{Trimodal Comparison}\label{ch3.4.4}
 
\paragraph{CRUX achieves the best external performance} CRUX achieves an external C-index of $\mathbf{0.761}$, the highest among all trimodal configurations and all methods evaluated in this study. This result surpasses the simple concatenation baseline (0.749), the text-anchored bag~\citep{kim2024miccai} (0.711), and SurvPath-Tri (0.698). Notably, CRUX's external performance also exceeds the best bimodal results---C+T CMTA (0.757) and P+T CMTA (0.758)---demonstrating that the text hub mechanism successfully extracts complementary information from both imaging modalities without suffering the generalization degradation observed in other trimodal methods.
 
\paragraph{The text hub resolves the trimodal overfitting problem} A consistent pattern across the baseline trimodal methods is that interaction-based approaches achieve high internal C-indices ($>0.80$) but degrade substantially on external validation, performing below both the text-only baseline and the best bimodal configurations. This pattern indicates that cross-modal interactions learned on the development cohort propagate domain-specific artifacts across modalities, amplifying overfitting. CRUX breaks this pattern: by routing all cross-modal exchange through the compact text semantic space and broadcasting the result back to imaging tokens via residual connections, the text hub preserves the institution-invariant anchoring while enabling effective information exchange.
 
\begin{figure*}[t]
\centering
\includegraphics[width=\textwidth]{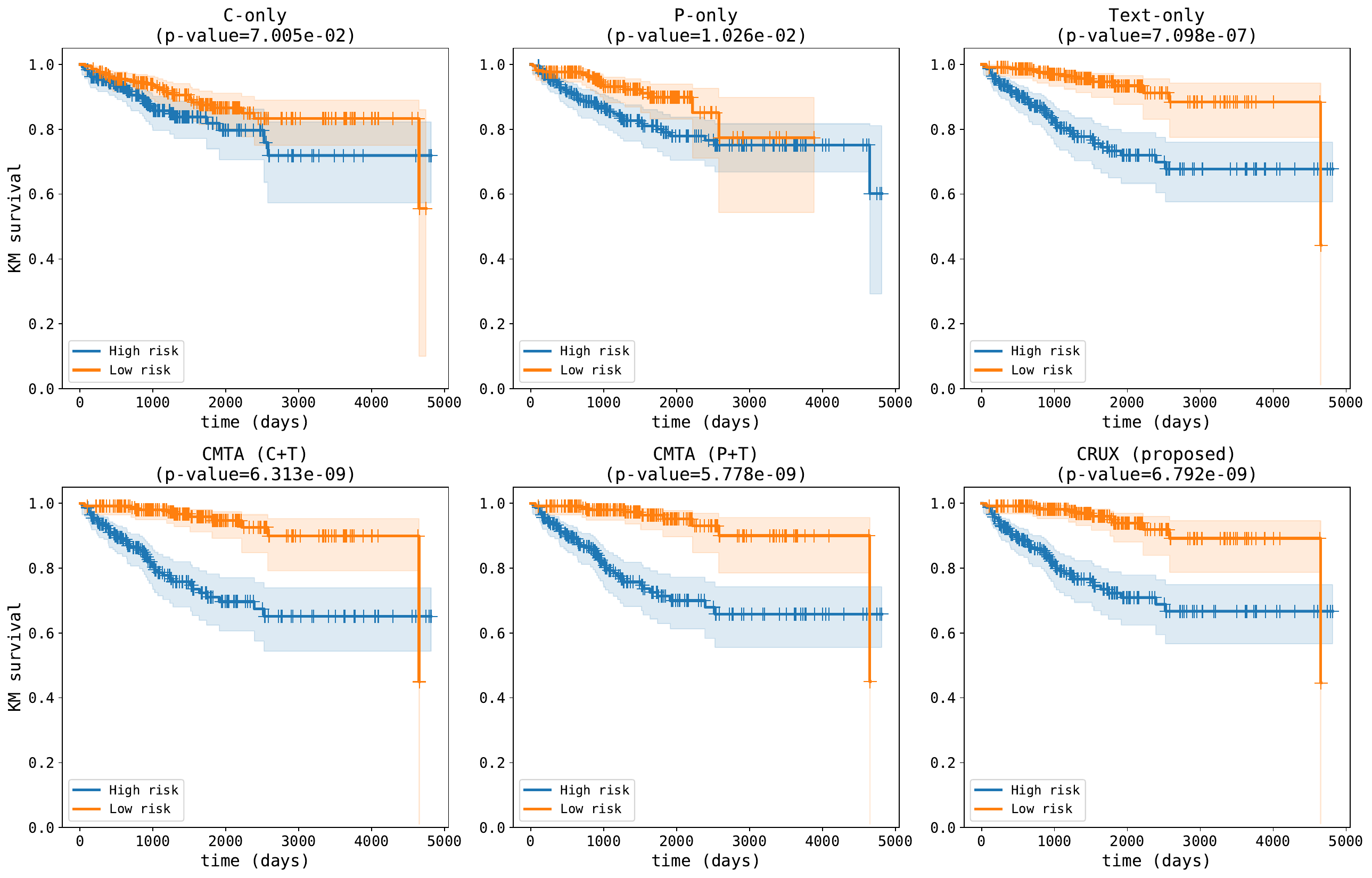}
\caption{Kaplan-Meier survival curves on the external validation cohort (PNUH). Patients are stratified into high-risk and low-risk groups based on the median predicted risk score. Top row: unimodal baselines. Bottom row: text-inclusive fusion methods. $p$-values are computed using the log-rank test.}
\label{fig:crux_km}
\end{figure*}

\paragraph{Risk stratification} Figure~\ref{fig:crux_km} presents Kaplan-Meier survival curves on the external validation cohort for six representative methods. CT-only fails to achieve statistically significant risk stratification ($p = 0.070$), while pathology-only reaches marginal significance ($p = 0.010$). Text-only achieves strong separation ($p = 7.1 \times 10^{-7}$), confirming its prognostic utility as a standalone modality. All three text-inclusive fusion methods---CMTA (C+T), CMTA (P+T), and CRUX---achieve highly significant stratification ($p < 10^{-8}$), with CRUX showing the clearest visual separation between high-risk and low-risk groups.
 
\subsection{Ablation Studies}\label{ch3.4.5}
 
\paragraph{Staging exclusion} As described in Section~\ref{ch3.2.1}, TNM staging and histological subtype are excluded from the text input to avoid information leakage, since these variables are derived from the same imaging modalities that the model is learning to interpret. Table~\ref{tab:crux_staging} validates this design decision using bimodal configurations that isolate the effect of text content. The text-only model shows negligible degradation when staging is removed (0.734 $\to$ 0.732, $\Delta = -0.002$), confirming that the four remaining non-staging variables---when encoded by BioLord---carry sufficient prognostic information for anchoring. C+T and P+T fusion are similarly robust ($\Delta = -0.007$ and $-0.003$, respectively), confirming that the non-staging text representation retains sufficient anchoring capacity for the proposed framework.
 
\begin{table}[t]
\centering
\caption{Effect of staging exclusion on external C-index. Bimodal configurations are used to isolate the effect of text content on each imaging modality independently. Staging: full clinical variables including TNM. Non-staging: demographics and tumor location only.}
\label{tab:crux_staging}
\resizebox{\columnwidth}{!}{\begin{tabular}{lccc}
\toprule
Method & With staging & Without staging & $\Delta$ \\
\midrule
Text only (TransMIL)  & 0.734 & 0.732 & $-$0.002 \\
C+T CMTA              & 0.764 & 0.757 & $-$0.007 \\
P+T CMTA              & 0.761 & 0.758 & $-$0.003 \\
\bottomrule
\end{tabular}
}
\end{table}

\paragraph{Cross-modal interaction strategy}
Table~\ref{tab:crux_ablation} compares different strategies for enabling CT--pathology interaction within the text-anchored framework.
Starting from the prior text-anchored bag~\citep{kim2024miccai}, which relies on implicit interaction through joint self-attention over all tokens (0.711),
the ablation evaluates three explicit cross-modal interaction designs.

Direct image-space interaction, where $\mathbf{z}_C$ and $\mathbf{z}_P$ attend to each other directly, achieves 0.705—slightly below the implicit baseline,
confirming that direct cross-attention between heterogeneous imaging modalities propagates domain-specific artifacts and is not effective in this setting.

Asymmetric text-mediated interaction, where $\mathbf{z}_C$ attends to $\mathbf{TQ}_P$ (the text-query summary of the opposite modality), improves to 0.723,
demonstrating that routing cross-modal information through text representations provides a more stable interaction pathway than direct image-space attention.
However, in this setting, the text representation remains static and does not incorporate information from other modalities, limiting its ability to enforce consistent integration.

The proposed text hub design, where $\mathbf{z}_C$ attends to $\hat{\mathbf{TQ}}_C$ (its own text-query summary enriched with cross-modal information aggregated from all modalities),
achieves the highest performance at 0.761.
Unlike asymmetric interaction, the text hub aggregates cross-modal information into a shared text representation and redistributes it back to each modality,
enabling globally consistent interaction.

\begin{sloppypar}
The clear hierarchy—hub (0.761) $>$ asymmetric (0.723) $>$ bag (0.711) $>$ direct (0.705)—demonstrates that cross-modal interaction is most effective when it is not only mediated by text representations,
but also structured through aggregation and redistribution, rather than relying on direct image-space attention or implicit joint aggregation.
\end{sloppypar}
 
\begin{table*}[t]
\centering
\caption{Comparison of cross-modal interaction strategies on external C-index. All variants share the same text-anchored encoding; they differ in how cross-modal information is exchanged, aggregated, and redistributed.}
\label{tab:crux_ablation}

\small
\resizebox{\textwidth}{!}{%
\begin{tabular}{lllc}
\toprule
Strategy & Cross-modal path & Text role & External \\
\midrule
\multicolumn{4}{l}{\textit{Implicit interaction}} \\
\quad Text-anchored bag (prior work)
  & $[\mathbf{z}_C; \mathbf{z}_P; \mathbf{TQ}_C; \mathbf{TQ}_P] \to \text{SA}$
  & Mixed tokens (no explicit aggregation)
  & 0.711 \\
\midrule
\multicolumn{4}{l}{\textit{Explicit interaction}} \\
\quad Direct image
  & $\mathbf{z}_C \leftrightarrow \mathbf{z}_P$
  & None
  & 0.705 \\
\quad Asymmetric text
  & $\mathbf{z}_C \leftrightarrow \mathbf{TQ}_P$
  & Static reference (no aggregation)
  & 0.723 \\
\quad \textbf{CRUX (proposed)}
  & $\mathbf{z}_C \leftrightarrow \hat{\mathbf{TQ}}_C$
  & Shared memory (aggregated \& broadcast)
  & \textbf{0.761} \\
\bottomrule
\end{tabular}
}

\end{table*}

\paragraph{Anchor modality choice}
A natural question is whether clinical text is essential as the hub modality, or whether CT or pathology could serve equally well. This is addressed by keeping the architecture fixed and changing only which modality acts as the hub---routing cross-modal interaction through CT or through pathology instead of through text---and retraining each variant under the same protocol. Table~\ref{tab:crux_anchor} reports the external C-index together with a model-agnostic measure of how domain-invariant each hub representation is: the AUC of a classifier trained to tell the external hospital apart from the development hospitals using the pooled hub representation (0.5 means the two are indistinguishable, i.e., institution-invariant; 1.0 means perfectly separable), and the corresponding maximum mean discrepancy (MMD). Using clinical text as the hub, which is the proposed CRUX configuration, gives the best external generalization (0.761); changing the hub to CT or to pathology degrades it to 0.732 and 0.680, respectively. The domain-invariance measure explains this ordering: the pathology hub is \emph{fully confounded} with hospital identity (domain-AUC 1.00, MMD 0.110), so its institution-specific artifacts propagate through the hub to every modality, whereas the text hub is the most invariant (0.542). The requirement for an effective anchor is therefore \emph{distributional stability across institutions} rather than raw predictive power, which is why clinical text---although it is the weakest stand-alone imaging-free signal---is the most reliable hub.

\begin{table}[t]
\centering
\caption{Anchor-modality (hub) ablation. The architecture is fixed and only the hub modality is changed (text, CT, or pathology), with each variant retrained under the same protocol; the text-hub row is the proposed CRUX configuration (Table~\ref{tab:crux_main}). Domain-AUC is the AUC of a hospital classifier on the pooled hub representation (0.5 = institution-invariant, 1.0 = perfectly separable); MMD is the maximum mean discrepancy. Best per column in bold.}
\label{tab:crux_anchor}
\resizebox{\columnwidth}{!}{\begin{tabular}{lccc}
\toprule
Hub (anchor) & External C-index & Domain-AUC $\downarrow$ & MMD $\downarrow$ \\
\midrule
Clinical text (CRUX) & \textbf{0.761} & \textbf{0.542} & 0.003 \\
CT                   & 0.732 & 0.726 & 0.004 \\
WSI (pathology)      & 0.680 & 1.000 & 0.110 \\
\bottomrule
\end{tabular}
}
\end{table}

\subsection{Attention Observation}\label{ch3.4.6}

Because CRUX routes cross-modal interaction through text summaries, its attention maps are read differently from conventional co-attention. The pre-hub maps show the evidence read by the CT- or pathology-side class token ($T_C$, $T_P$), and the post-hub maps show the same read after enrichment by the text hub ($\hat{T}_C$, $\hat{T}_P$). These are qualitative observations of internal evidence routing, not tumor localization or causal attribution of the final risk.

\begin{figure*}[p]
\centering
\subfloat[]{\includegraphics[width=0.5\textwidth]{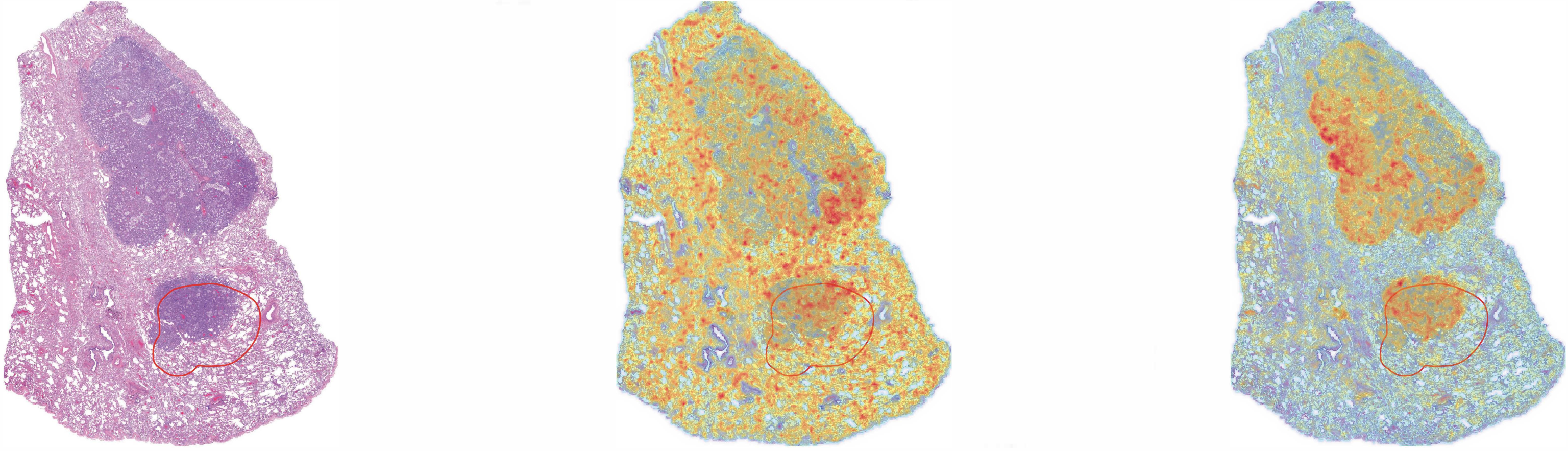}}\\[-12pt]
\subfloat[]{\includegraphics[width=0.5\textwidth]{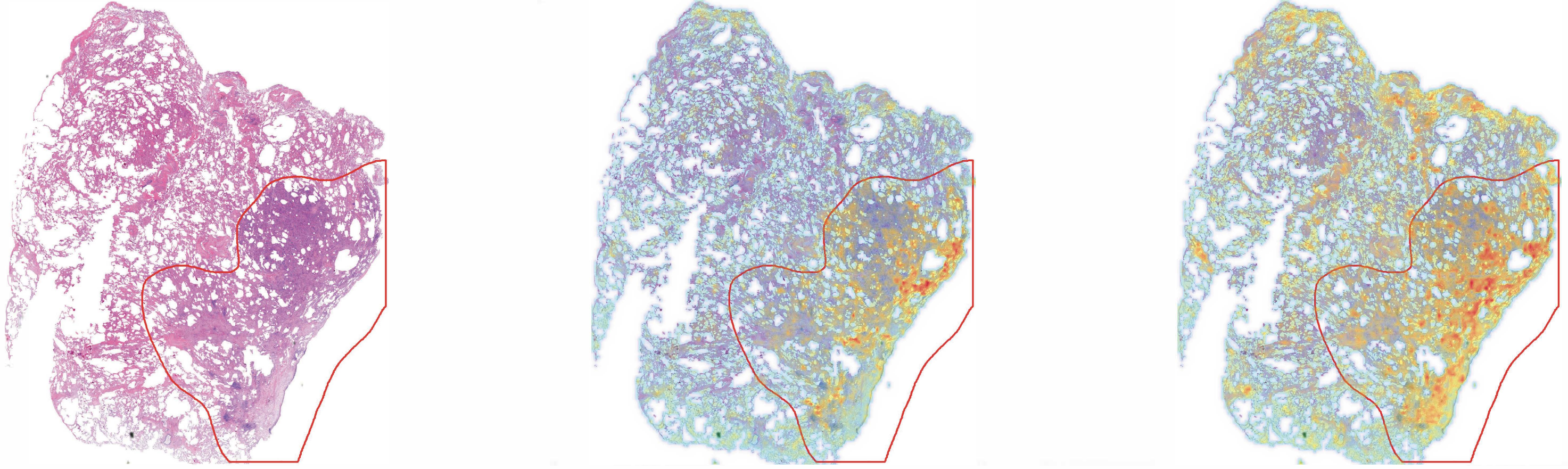}}\\[-12pt]
\subfloat[]{\includegraphics[width=0.5\textwidth]{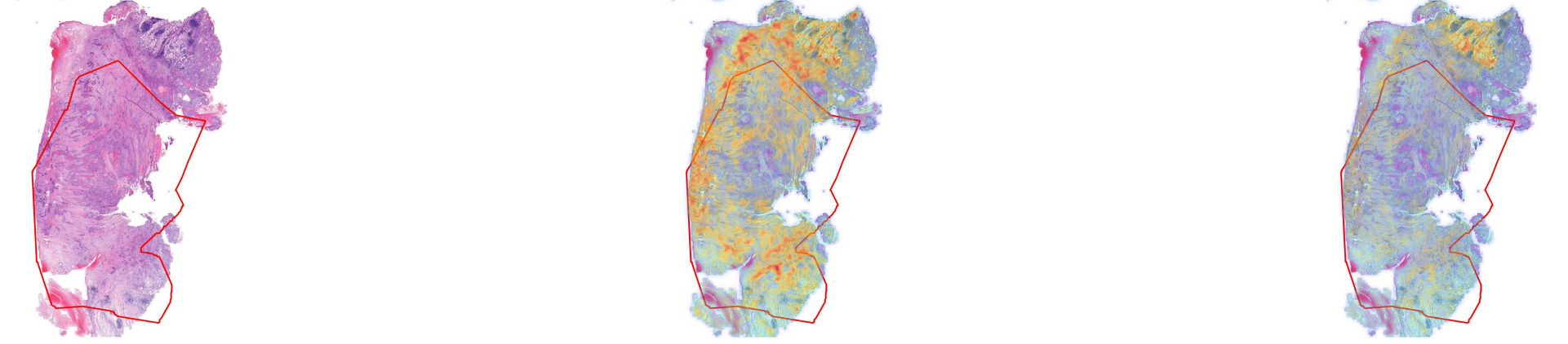}}\\[-12pt]
\subfloat[]{\includegraphics[width=0.5\textwidth]{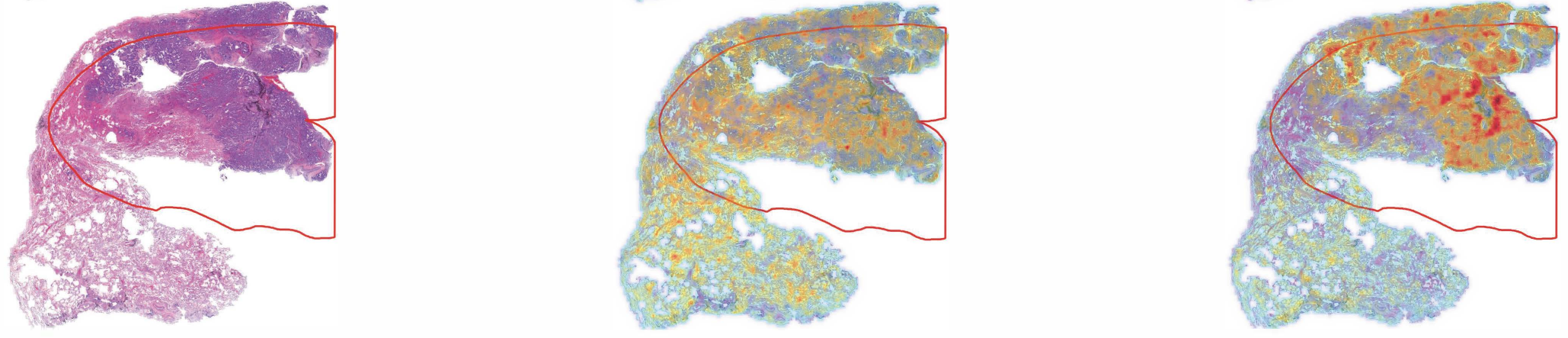}}\\[-12pt]
\subfloat[]{\includegraphics[width=0.5\textwidth]{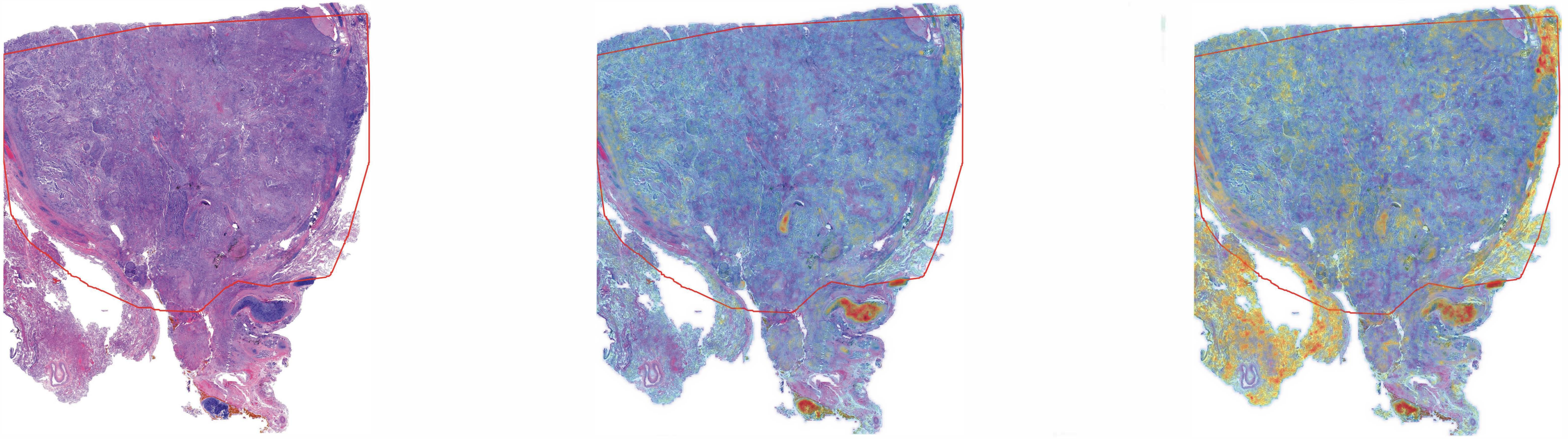}}
\caption{Hub-mediated WSI attention observation in CRUX for five PNUH resection cases. Each panel shows the H\&E image with a pathologist-marked tumor outline (red; reference only), the pre-hub pathology-side class-token attention from $T_P$ to WSI patches, and the post-hub rollout from $\hat{T}_P$ to the WSI patch space. Red/orange indicates higher attention and blue/purple lower attention. The maps are qualitative evidence-routing patterns before and after text-hub exchange, not tumor localization or causal attribution of the final risk.}
\label{fig:crux_attn_wsi}
\end{figure*}

\begin{figure*}[p]
\centering
\subfloat[]{\includegraphics[width=0.8\textwidth]{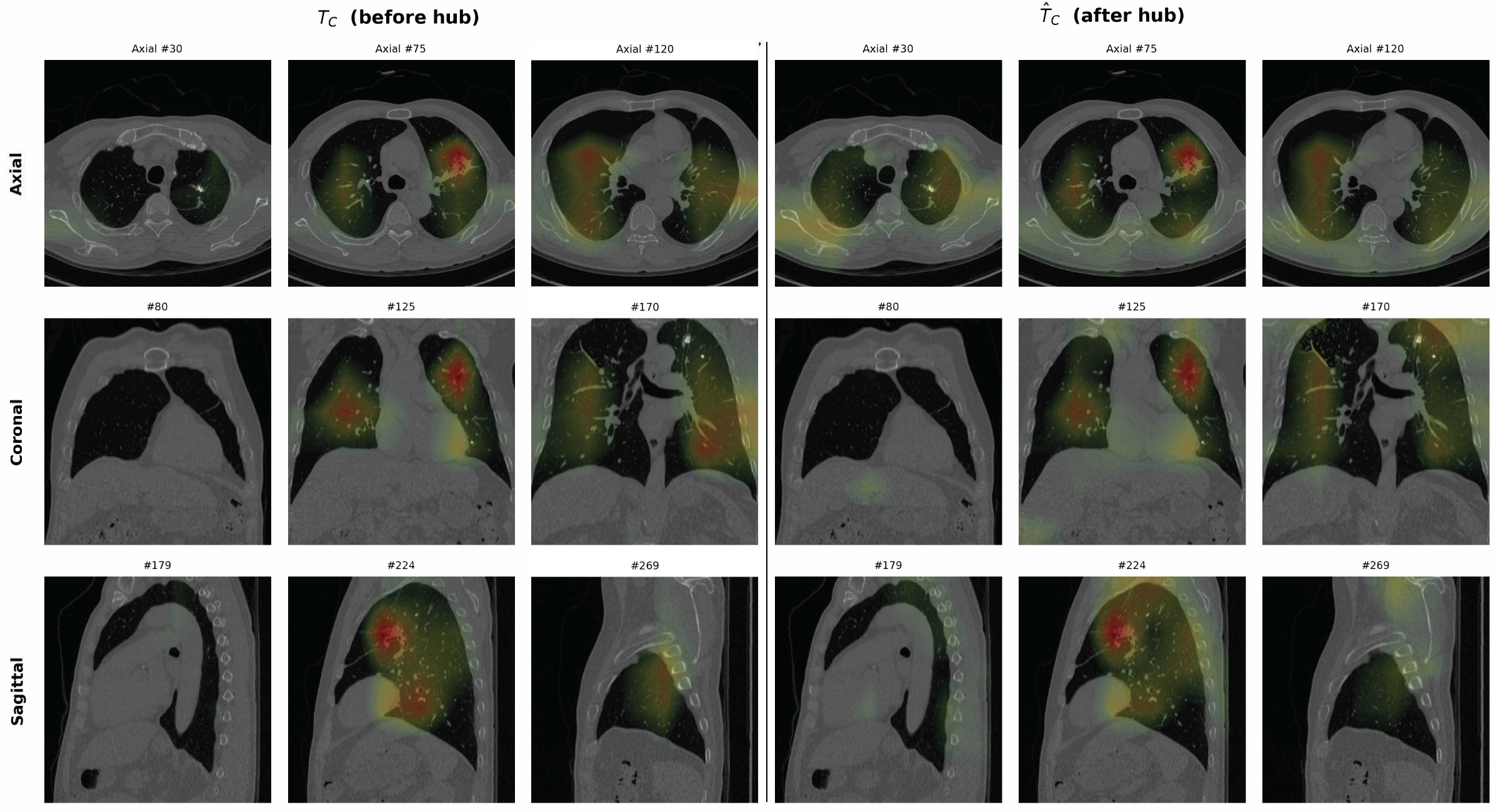}}\\
\subfloat[]{\includegraphics[width=0.8\textwidth]{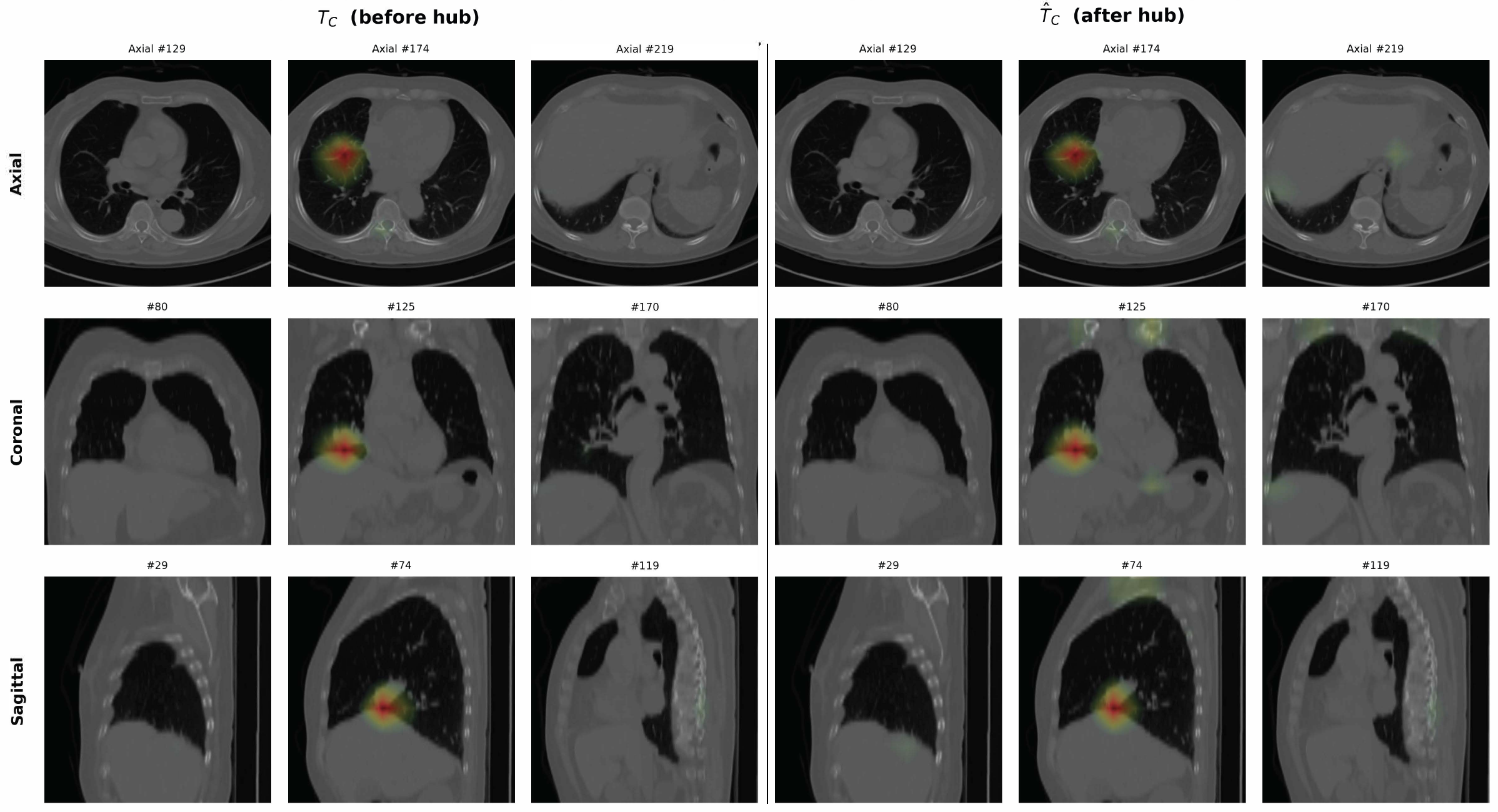}}
\caption{Hub-mediated CT attention observation in CRUX for the two cases (a, b) that also appear in Figure~\ref{fig:crux_attn_wsi}. CT maps are shown in axial, coronal, and sagittal views using coarse $6\times6\times6$ CT tokens, with slices centered on the pre-hub $T_C$ peak, comparing the pre-hub ($T_C$) and post-hub ($\hat{T}_C$) evidence reads. Red/orange indicates higher attention. These are macro-scale qualitative observations, not voxel-level lesion masks.}
\label{fig:crux_attn_ct}
\end{figure*}

Figure~\ref{fig:crux_attn_wsi}(a--e) indicates that the hub does not act uniformly on WSI evidence: it is largely \emph{preserved} when already consistent with the hub context (b), \emph{refined} from a diffuse response toward a smaller subset of regions (a, d), or \emph{suppressed and redistributed} when a pathology-only salient region is less consistent with the cross-modal context (c, e); the red tumor outlines serve as anatomical references only. The CT maps in Figure~\ref{fig:crux_attn_ct}(a--b), shown for two of the same cases, follow the same pattern at a coarser ($6\times6\times6$) macro-anatomical scale: the lesion-side response is preserved while diffuse responses are pruned (a), and a focal CT response can remain stable even when the paired WSI map is redistributed more diffusely (b versus Figure~\ref{fig:crux_attn_wsi}(e)), reflecting the different spatial scales of the two modalities.

Overall, the text hub \emph{recontextualizes} modality-specific image evidence---preserving, refining, suppressing, or redistributing it---rather than amplifying it or enforcing CT--WSI spatial correspondence, consistent with the design of CRUX as a text-mediated routing framework rather than an image-space alignment model.

\section{Discussion and Conclusions} \label{ch3.5}
 
This paper presented CRUX, a text-anchored cross-modal fusion framework for survival prediction from CT, pathology, and clinical text. The central idea is to encode clinical variables through biomedical language models and use the resulting institution-invariant embeddings as semantic axes for organizing imaging features and mediating cross-modal interaction. Three design principles underlie the framework: (1) language model--encoded text provides richer and more generalizable representations than scalar tabular features, (2) text-anchored encoding aligns heterogeneous imaging modalities through a shared semantic reference frame, and (3) a text-mediated hub enables CT--pathology interaction without direct image-space cross-attention that would propagate domain-specific artifacts.
 
Experiments on a multi-institutional lung cancer cohort with external validation confirmed these principles. Text-only prediction using BioLord embeddings of four non-staging variables achieved an external C-index of 0.732, substantially outperforming CoxPH applied to the same variables as scalars (0.684) and demonstrating that language model encoding captures implicit prognostic knowledge beyond what scalar features convey. Text inclusion systematically improved cross-institutional robustness across all fusion architectures: all CT+Text configurations exceeded 0.74 externally, and P+T CMTA reached 0.758. Most importantly, CRUX achieved the highest external C-index of 0.761 among all configurations evaluated, surpassing both the simple concatenation baseline (0.749) and the best bimodal results. The ablation study revealed a clear hierarchy among cross-modal interaction strategies---hub (0.761) $>$ asymmetric (0.723) $>$ bag (0.711) $>$ direct (0.705)---confirming that cross-modal interaction is most effective when mediated through text representations enriched by a dedicated exchange mechanism.
 
\paragraph{Limitations} Several limitations of this work merit discussion.
 
First, CRUX assumes that all three modalities---CT, pathology, and clinical text---are available for every patient. This assumption is satisfied in the curated multi-institutional dataset used in this study, but it does not hold universally in clinical practice. Pathological specimens may be unavailable for patients who do not undergo surgical resection, and retrospective cohorts frequently have incomplete imaging archives. When an entire imaging modality is absent, CRUX's text-anchored encoding cannot be applied for that modality, and the framework provides no mechanism for cross-modal recovery. Extending the framework to entirely missing modalities---via cross-modal recovery rather than assuming complete availability---is an important direction for future work.
 
Second, the clinical text used in this framework is not an independent information source: the recorded variables are themselves derived from imaging assessment and clinical observation. The text modality therefore serves as a \emph{structured semantic prior} that regularizes the fusion process, rather than as a carrier of novel prognostic information. While this regularization role is empirically effective---as evidenced by the consistent improvement in external generalization when text is included---it also implies a ceiling on the benefit that text anchoring can provide. The framework cannot compensate for information that is absent from both imaging modalities and the clinical record.
 
Third, the pathology domain gap remains a fundamental challenge that text anchoring mitigates but does not resolve. Pathology features exhibited the largest institution-specific variation among the three modalities, and all models involving pathology showed substantial generalization gaps on the external cohort. This observation aligns with known challenges in computational pathology, where staining protocols, tissue preparation, and scanning equipment vary considerably across institutions. Addressing this limitation would require domain adaptation techniques at the feature extractor level, which is beyond the scope of the text-anchored fusion approach.
 
Finally, this study evaluates a single cancer type (non-small cell lung cancer) using data from South Korean hospitals. The generalizability of the text-anchored approach to other cancer types, clinical endpoints, and demographic populations remains to be established.
 
\paragraph{Outlook} CRUX demonstrates that effective multimodal fusion can be achieved when all modalities are present, using text as a semantic anchor to bridge heterogeneous imaging spaces and to stabilize fusion across institutions. Clinical utility, however, requires robustness not only to distributional shift but also to the absence of entire modalities; extending text-anchored fusion with an explicit mechanism for recovering missing modalities is a natural next step toward deployable multimodal prognosis.

\section*{CRediT authorship contribution statement}
\textbf{Kyungwon Kim:} Conceptualization, Methodology, Software, Formal analysis, Investigation, Visualization, Writing -- original draft, Writing -- review \& editing.
\textbf{Dosik Hwang:} Supervision, Funding acquisition, Resources, Project administration, Writing -- review \& editing.

\section*{Ethics statement}
This study used only publicly available, de-identified data provided by the Open AI Dataset Project (AI-Hub, Republic of Korea); no new data were collected from human participants by the authors.

\section*{Declaration of competing interest}
The authors declare that they have no known competing financial interests or personal relationships that could have appeared to influence the work reported in this paper.

\section*{Data availability}
The lung cancer dataset used in this study is provided by the Open AI Dataset Project (AI-Hub, Republic of Korea) and is available at \url{https://www.aihub.or.kr/aihubdata/data/view.do?currMenu=115&topMenu=100&dataSetSn=71394}. Access requires an AI-Hub account and, under the repository's data-use policy, is currently restricted to nationals of the Republic of Korea.

\bibliographystyle{elsarticle-harv}
\bibliography{references}

\end{document}